# RESEARCH

Open Access

# Microbial community pattern detection in human body habitats via ensemble clustering framework

Peng Yang[1], Xiaoquan Su[2], Le Ou-Yang[3], Hon-Nian Chua[1], Xiao-Li Li[1], Kang Ning[2*]

**Abstract**

**Background:** The human habitat is a host where microbial species evolve, function, and continue to evolve. Elucidating how microbial communities respond to human habitats is a fundamental and critical task, as establishing baselines of human microbiome is essential in understanding its role in human disease and health. Recent studies on healthy human microbiome focus on particular body habitats, assuming that microbiome develop similar structural patterns to perform similar ecosystem function under same environmental conditions. However, current studies usually overlook a complex and interconnected landscape of human microbiome and limit the ability in particular body habitats with learning models of specific criterion. Therefore, these methods could not capture the real-world underlying microbial patterns effectively.

**Results:** To obtain a comprehensive view, we propose a novel ensemble clustering framework to mine the structure of microbial community pattern on large-scale metagenomic data. Particularly, we first build a microbial similarity network via integrating 1920 metagenomic samples from three body habitats of healthy adults. Then a novel symmetric Nonnegative Matrix Factorization (NMF) based ensemble model is proposed and applied onto the network to detect clustering pattern. Extensive experiments are conducted to evaluate the effectiveness of our model on deriving microbial community with respect to body habitat and host gender. From clustering results, we observed that body habitat exhibits a strong bound but non-unique microbial structural pattern. Meanwhile, human microbiome reveals different degree of structural variations over body habitat and host gender.

**Conclusions:** In summary, our ensemble clustering framework could efficiently explore integrated clustering results to accurately identify microbial communities, and provide a comprehensive view for a set of microbial communities. The clustering results indicate that structure of human microbiome is varied systematically across body habitats and host genders. Such trends depict an integrated biography of microbial communities, which offer a new insight towards uncovering pathogenic model of human microbiome.

## Background

### Metagenomic background

The human body is a content that complex microbial communities are living inside and on. This microbiome occupies body habitats and endows us with ecosystem functions, such as nutrition, pathogen resistance and immune system development [1,2], to help maintain our health. Hence systematically defining the "normal" states of human microbiome is an important step towards understanding role of microbiota in pathogenesis [3]. However, the majority of microbiomes have been poorly investigated.

To understand the principle of human microbiome, prior research concentrated on particular body habitats [3-8]. For example, Turnbaugh et al. [9] investigated the gut microbiome in obese and lean twins to address how host, environmental condition and diet influence the

* Correspondence: ningkang@qibebt.ac.cn
[2]Computational Biology Group of Single Cell Center, Shandong Key Laboratory of Energy Genetics and CAS Key Laboratory of Biofuels, Qingdao Institute of Bioenergy and Bioprocess Technology, Chinese Academy of Science, Qingdao 266101, China
Full list of author information is available at the end of the article microbial components. Grice et al. [10] targeted human skin microbiome to characterize its topological and personal variations within multiple sites. Bik et al. [11]'s research indicated the distinctness of microbial structure on oral cavity and tongue.

However, human microbial habitats are not isolated with one another; instead they reveal community structure correlation across body habitats [12]. In this case, ensemble of different habitat samples could bring global and full-scale insights into microbiome. Recent studies had aggregated microbial samples from different body habitats to perform a comprehensive study. Costello et al. [13] surveyed the microbiomes that were gathered from 27 body habitats of nine adults. Mitreva [12] carried out the extensive sampling on 18 body habitats from 242 individuals. In order to establish a global insight of human microbiome, they built a "whole-body" microbial similarity network where the nodes were consisted of metagenomic samples from multiple human body sites and the edges as pair-wise phylogenetic similarity of samples were measured in terms of their shared evolutionary history. Clustering approaches [14] had been applied on this large-scale similarity network to group samples that shared more similar phylogenetic structures with each other within the clusters than other ones. From these clusters, researchers could infer how microbial patterns were affected by body habitat, host gender and environmental condition with time. Costello et al. [13] proposed a hierarchical clustering algorithm on a microbial community network and found out personal microbiota relatively stable within habitats over time. Turnbaugh et al. [9] identified two distinct functional modules on gut microbiome via principal components analysis (PCA) and hierarchical clustering algorithm, and experimental results disclosed that microbiome within same clusters carried out similar ecosystem-level functions. Mitreva [12] adopted a centroid-based clustering algorithm and discovered the co-variation and co-exclusion of microbiome between different habitats.

### Current Limitations

Clustering approach aims to group metagenomic samples with similar phylogenetic patterns. It can be achieved by various algorithms that differ significantly in terms of computational principles and measures, by which each generated clustering results can be viewed as taking a different "look" through data (as shown in Table 1). However, most of prior studies employ one particular clustering approach, by which the clustering outputs tend to be specific towards the criterion of the proposed approach. For example, density-based clustering algorithm groups samples that are densely connected in similarity network. However, true microbial communities are not limited to densely connected structures; samples with sparsely microbial structure widely exist in the lake [15]. Graph partition-based clustering such as MCL [16] and K-means clustering [17] explores the best partition of a network. But these algorithms do not allow the overlaps between clusters. Therefore, they are unable to discover shared microbe between two communities, such as some species that could adapt in multi-environmental conditions like microbial mats and biofilms. Hierarchical clustering algorithm [18] learns the hierarchical structure of a network, which has been used in [13], but hierarchical structure is determined by local optimization criterion as such there is no global objective function, which might lead to small clusters with only part of similar samples. Distribution-based clustering approach, like expectation-maximization (EM) [19], identifies the clusters that follow statistical Condorcet criteria. But statistical model for microbial community remains rarely known and therefore it is difficult to evaluate reliability of the results.

### Advantage of proposed Ensemble clustering framework

Ideally, a clustering algorithm should be able to exploit clustering patterns as comprehensive as possible. However, as we have mentioned above, few algorithms are capable of taking into consideration all factors. Different clustering algorithms may produce different partitions of the network. Given multiple clustering results, we need to explore their information and output more robust results that can exploit the complementary nature of these patterns.

Ensemble clustering was proposed recently which has been successfully used to solve many community detection problems [20-23]. Thus, we use ensemble clustering framework to integrate the various kinds of clusters (here we call them base clustering results) and output more comprehensive results. In this study, we first construct a consensus matrix which measures similarity of samples based on co-occurrence of samples in base clustering results [24]. Next we apply Symmetric Nonnegative Matrix Factorization (NMF) [25] on the consensus matrix to derive clusters. Symmetric NMF provides a lower rank approximation of a nonnegative matrix, which could be easily related to the clustering of the nonnegative data. As mentioned in [25], the factorization of the consensus matrix will generate a clustering assignment matrix that could capture the cluster structure inherent in the network.

Unlike prior researches that applied single cluster algorithm on particular habitat microbiome, our framework assembled clustering algorithms of different human microbiome in different body habitats. We carried out our experiments to demonstrate its capability in capturing the microbial community. Experimental results showed that predicted clusters were capable of revealing the spatial and gender roles of human microbiota and eventually elaborated human microbiome biogeography, which provided new insights about disease pathogenesis of human microbiome [9,12,13].

Table 1 Summary of four particular clustering approaches

| Clustering Approaches | Characteristics | Limitations on microbial pattern |
|---|---|---|
| Density-based clustering | Clusters are defined as connected dense regions in the network | True microbial community are not limited to densely connected structures; sparsely **microbial structure** still exists |
| Graph partition-based clustering | Clusters are generated via graph partitioning techniques | Partition based algorithms do not allow the overlaps between clusters. Therefore, they are unable to discover shared microbe among clusters, such as some species that could adapt in multi-environmental conditions like microbial mats and biofilms |
| Hierarchical clustering | Clusters are built based on an agglomerative clustering model that shows relations between the members and groups | Hierarchical structure is determined by local optimization criterion as such there is no global objective function, which might lead to small clusters with only part of similar samples |
| Distribution-based clustering | Clusters are modelled using statistical distributions | Statistical models of microbial communities are still unknown and need to be further explored |

## Material and methods

In this section, we first briefly introduced the experimental data, the similarity measurements of metagenomic samples and GPU based fast similarity matrix computing. Then we described the schema of ensemble clustering framework and its phases to structure microbial community.

### Experimental data

In this work, we used 1920 metagenomic samples from the project "Moving pictures of human microbiome" [26] to build the microbial matrix and similarity network (refer to section **"Similarity measurements of metagenomic samples"** for details). A sample of metagenomic matrix and network were illustrated in Figure 1 and the similarity matrices of all datasets were shown in Additional file 2: Table S1. GPU-Meta-Storms [27] were performed to measure structural similarity of metagenomic samples (Efficiency of GPU-Meta-Storms is shown in Additional file 2: Figure S1). Metagenomic samples were annotated by two **meta-labels**: Habitat = {gut, skin, oral cavity} defined human body habitat the samples live in, while Gender = {male, female} defined the gender of host the samples inhabit. Combining the two **meta-labels**, each sample was partition into one of six meta-classes, they were {male & gut, male & skin, male & oral cavity, female & gut, female & skin, female & oral cavity}. Table 2 summarized the distribution of 1920 metagenomic samples on three body habitats and two host genders.

### Similarity measurements of metagenomic samples

The scoring function of Meta-Storms [27] compared two microbial samples' structure by calculating the maximum common component of their common phylogenetic tree considering the $\beta$-diversity, phylogenetic distance and abundance of each species (**Formula 1**). The scoring function first evaluated the *common abundance* of each species on the leaf node, which was considered as the smaller abundance value in two samples. These abundance values were propagated to their ancestors iteratively, and the accumulative common abundance values at the root node reflected the overall similarity between the two metagenomic samples, which could be computed using Similarity (Root) defined in **Formula 1**.

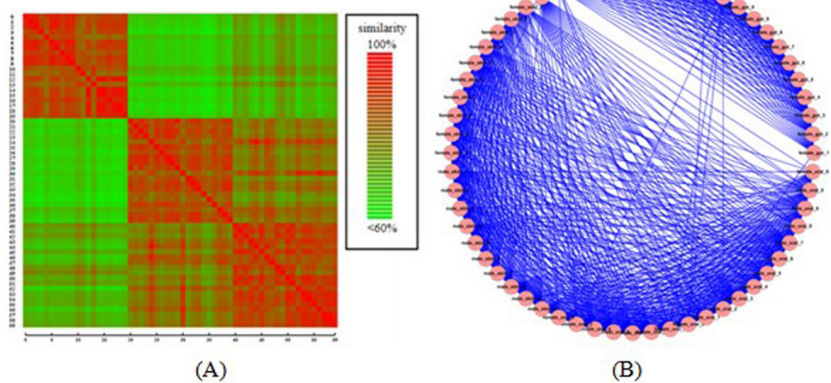

**Figure 1 An example of (A) similarity matrix and (B) its similarity network**. In the matrix, each tile indicates a similarity value between 2 samples by colour gradient from red (high) to green (low). In the network, each node represents a sample, and edges represent similarity values in the matrix.

**Table 2 1920 microbial samples on six human body habitats**

|        | Gut | Skin | Oral | Total |
|--------|-----|------|------|-------|
| Male   | 331 | 698  | 366  | 1395  |
| Female | 130 | 262  | 133  | 525   |
| Total  | 461 | 960  | 499  | 1920  |

$$\text{Similarity}(X) = \begin{cases} \text{Common Abundance}(X) & \text{If } X \text{ is a leaf node} \\ \text{Common Abundance}(X) \text{ If } X \text{ is an internal node} \\ +\text{Similarity}(X.\text{Left}) \\ +\text{Similarity}(X.\text{Right}) \end{cases} \quad (1)$$

Then we constructed the similarity matrix based on the pair-wised similarity among all sample pair (Figure 1 **(A)**). Exploiting the multi-core architecture of the GPU [28], Formula 1 could be invoked in parallel using a large number of threads to compute similarity between different pairs of metagenomic samples. To compute the pair-wise similarity matrix for $N$ samples, we spawned $N * N$ threads in the GPU such that each similarity value in the matrix was processed by an independent thread.

Figure 2 illustrated the GPU computing workflow: to build the common phylogenetic tree, we first loaded and initialize abundant specie data from the file system to main memory; this data was then reloaded to the GPU for computing. When all threads of the GPU kernel had been completed (Figure 1, step 3, the key step), these values were returned back to RAM to populate the similarity matrix, which was then stored in the file system.

**Ensemble clustering framework**

In this subsection, we proposed a novel ensemble clustering framework, namely Meta-EC, to perform microbial community pattern detection. The framework consisted of two stages: a generation phase where a consensus matrix was constructed based on base clustering results and an identification phase in which a symmetric NMF-based clustering was used to detect reliable clusters from the consensus matrix. The schema of our Meta-EC algorithm was presented in Figure 3.

**Terminology**: After computing the pair-wise similarity matrix of the metagenomic samples, we used it to construct the microbial similarity network that was reformatted as a simple undirected graph $G = (V, E)$, where $V$ defined a vertex set which containeed $|V| = N$ vertices, and $E$ an edge set. A vertex $v \in V$ represented a metagenomic sample and a weighted edge $e \in E$ represented the polygenetic structure similarity of two samples (Figure 1**(B)**). A cluster $C_i = (V_c^i, E_c^i)$ was a subnetwork of $G$ such that $V_c^i \subset V$ and $E_c^i$ was the set of edges induced by $V_c^i$ from $G$. A microbial community of $G$ was a set of predicted microbial clusters, defined as $\{C_1,...,C_m\}$.

**Generation phase**: When the similarity network was ready, a set of base clustering results were calculated by applying four clustering algorithms (base clustering algorithms) on the similarity network with different initializations, as shown in Figure 3(**A**). The base clustering algorithms included EM algorithm, K-mean clustering, hierarchical clustering and density-based clustering, as present in Table 1 and Additional file 1: Section 1. A consensus matrix $W$ was introduced to measure the co-occurrence of samples in clusters of the base clustering results. Each $W_{ij}$ indicated the number of base clustering results in which sample $i$ and sample $j$ were assigned to the same cluster, divided by the total number of base clustering results. Therefore, matrix $W$ took into consideration all generated clusters and reflected the co-clusters similarity between each pair of samples based on different clustering criterions. The higher the value of $W_{ij}$, the more likely sample $i$ and sample $j$ belonged to the same cluster.

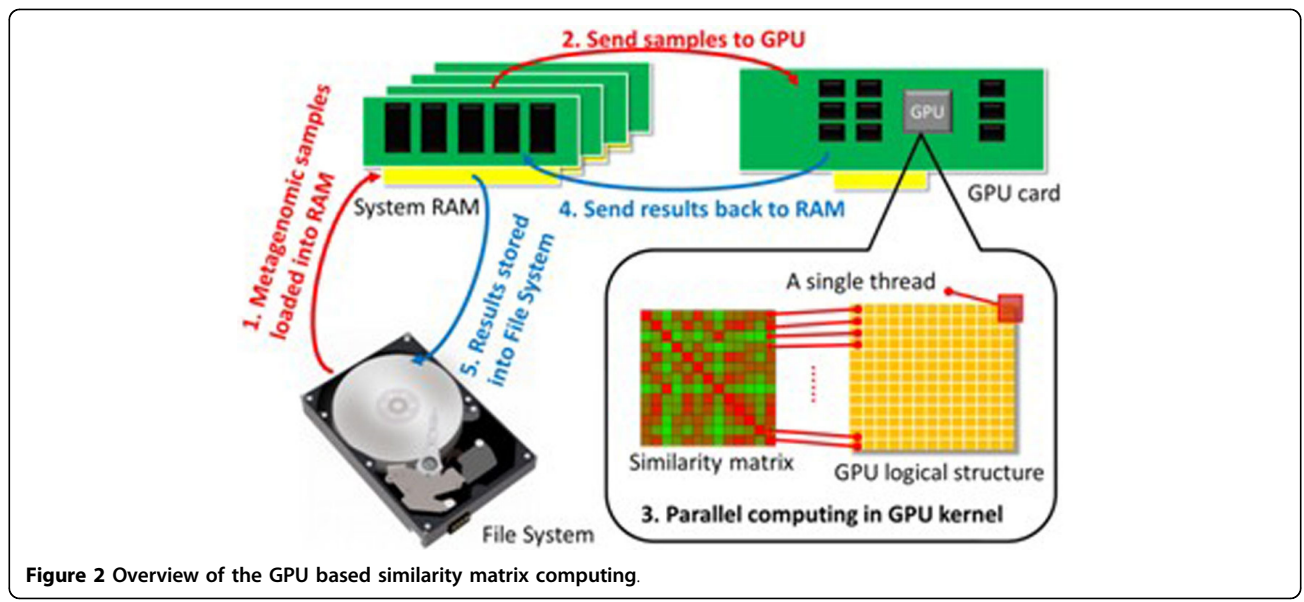

Figure 2 Overview of the GPU based similarity matrix computing.

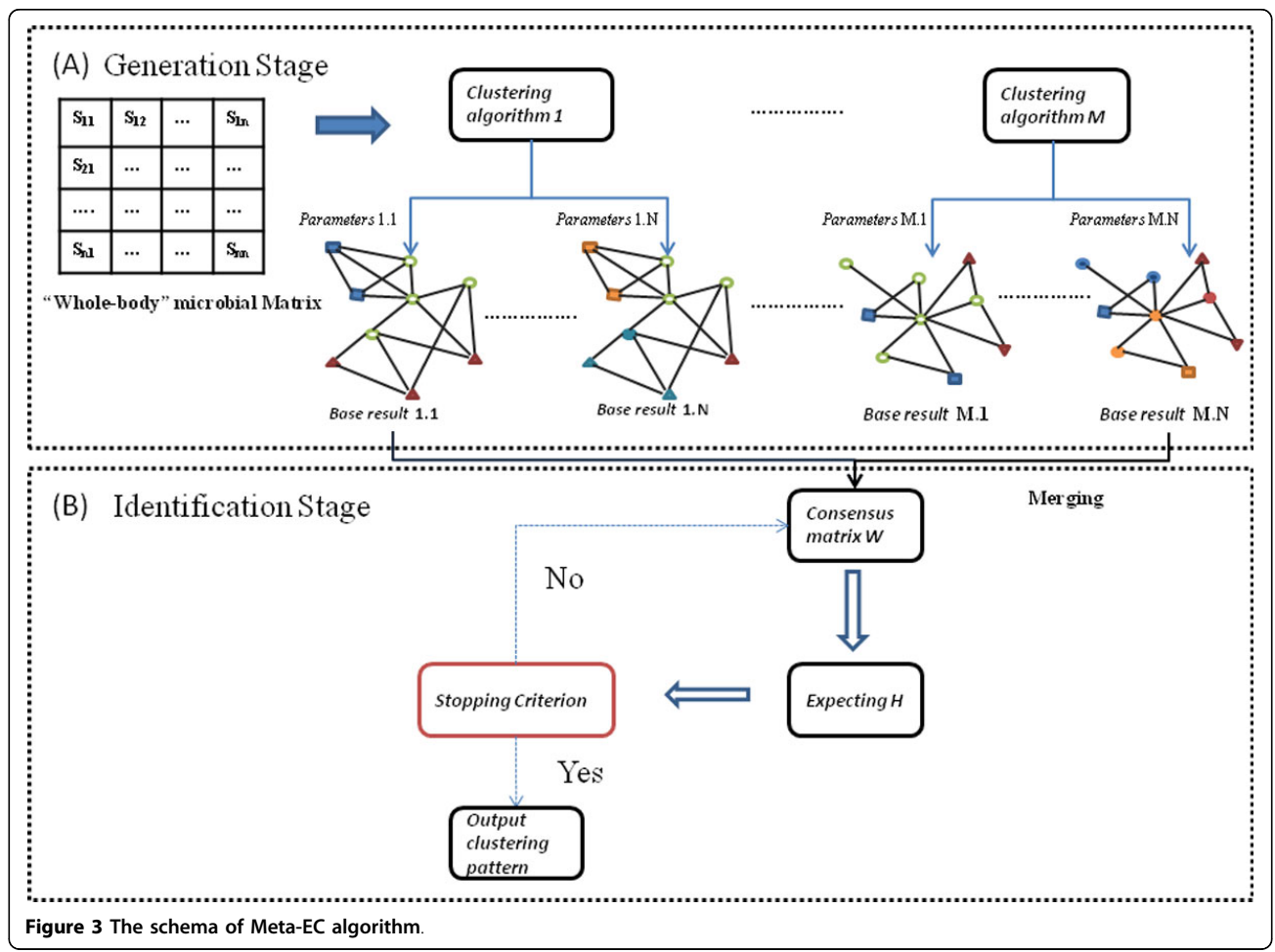

**Figure 3 The schema of Meta-EC algorithm.**

**Identification phase**: When the consensus matrix was constructed, we applied a symmetric NMF-based clustering algorithm on this matrix to derive the clusters. The flowchart of this algorithm was shown in Figure 3 (**B**). The main idea of this algorithm was outlined as follows:

The symmetric NMF defined in Equation (2) was suitable for network clustering based on similarity matrix $W$:

$$min_{H \geq 0} D\left(W | HH^T\right) \qquad (2)$$

Here $D(W|HH^T)$ was a predefined cost function $K$ and $K$ was the predefined number of clusters. $H$ was a cluster indicator matrix in which each entry $h_{i,k}$ denoted the real-valued membership of sample $i$ belonging to cluster $k$. So we could easily infer the clustering assignment of sample $i$ from the $i$-th row of $H$. In this study, we used Kullback-Leibler (KL) divergence [28] as the cost function, which could be represented as:

$$D\left(W|HH^T\right) = D_{KL}\left(W|HH^T\right) = \sum\nolimits_{i,j=1}^{N} W_{ij} \log\left(\frac{W_{ij}}{(HH^T)_{ij}}\right) - W_{ij} + (HH^T)_{ij} \qquad (3)$$

We chose KL-divergence as the cost function since it was free of noise parameter and had been widely used in NMF.

A sample may belong to more than one cluster, but it seldom belonged to all clusters. Thus, the cluster indicator matrix $H$ should be sparse. To achieve sparsity of the solution of $H$, a $L1$-norm regularization for $H$ was integrated. Neglecting constants and adding the $L1$-norm regularization for $H$, the modified formulation was as follows:

$$\min_{H \geq 0} \left( -\sum_{i=1}^{N} \sum_{j=1}^{N} \left( W_{i,j} \log(HH^T)_{i,j} - (HH^T)_{i,j} \right) + \sum_{i=1}^{N} \sum_{z=1}^{K} \beta h_{i,z} \right) \quad (4)$$

where the hyper-parameter $\beta > 0$ controlled the sparsity of $H$, and $H \geq 0$ was the cluster indicator matrix.

**Solution to NMF-based Ensemble Clustering:** Minimization of the cost function in equation (4) with constraints formed a constrained nonlinear optimization problem. Similar to [29,30], we adopted the multiplicative update rule [31] to estimate $H$, which was widely accepted as a useful algorithm in solving nonnegative matrix factorization problem. By the multiplicative update rule, we obtained the following update rules for $h_{i,z}$:

$$h_{i,z} \leftarrow \frac{h_{i,z}}{2} + \frac{1}{2} h_{i,z} \frac{\sum_{j=1}^{|V|} \frac{W_{i,j} h_{j,z}}{\sum_{l=1}^{K} h_{i,l} h_{j,l}}}{\sum_{j=1}^{|V|} h_{j,z} + 0.5\beta} \quad (5)$$

We iteratively updated $H$ according to the updating rule (5) until they satisfied a stopping criterion. Let $H_l$ be the cluster indicator matrix at iteration time $l$ ($l > 1$). The algorithm was stopped whenever $||H_l - H_{l-1}||_1 < \rho$, where $\rho$ was a predefined tolerance parameter. Here we set $\rho = 10^{-6}$ as the default value of tolerance parameter. In addition, the maximum of iteration time was limited to 200 iterations if the stopping criteria $\rho$ was unsatisfied. In order to avoid local minimum, for random initialization, we repeated the algorithm 10 times with random initial conditions and chose the results with lowest value of the cost function (4).

**From cluster indicator matrix to microbial clusters:** Similar to [32], we obtained microbial clusters from cluster indicator matrix $H$ by taking the threshold $\tau$ to assign a sample to a cluster when its weight for the cluster exceeded $\tau$. In this way, we obtained the resultant sample-cluster membership matrix $H^* = \left( h_{i,z}^* \right)$, where $h_{i,z}^* = 1$ if $h_{i,z} \geq \tau$ and $h_{i,z}^* = 0$ if $h_{i,z} < \tau$. Here, $h_{i,z}^* = 1$ mean sample $i$ was assigned to detected cluster $z$ and $h^*$ mean the final output of $h$. After completing these steps, we obtained the refined clusters $E_K$ that satisfied the following conditions:

$$E_K = \left\{ \{C_1, ..., C_K\} : v_i \in C_Z, if .h_{i,z}^* = 1 \right\} \quad (6)$$

where $i = 1,...,N$ and $z = 1,...,K$.

We summarized the whole algorithm in Figure 4.

## Results

In this section, we focused on evaluating the effectiveness of Meta-EC algorithm. Before presenting the experimental results, we first introduced our experiment design: evaluation metrics and experimental settings in

- Input:
- $W$: Consensus matrix that combines the networks from four base clustering results;
- $H_{initial}$: Initial value of $H$;
- $\beta$: Rate parameter of the coefficient of the sparse regularization term;
- $\rho$: Tolerance parameter for stopping criterion;
- $\tau$: Threshold parameter for obtaining sample-cluster membership matrix;
- Output:
- $E_K$: The final detected microbial cluster patterns.
- $J$: Value of the objective function
- Main algorithm:
1. Initializing the consensus matrix $W$ that combines clustering results from four different base algorithms;
2. Initializing the sample-cluster propensity matrix $H$;
3. Updating the $H$ according to rules (5);
4. Repeat the step 3 utile meeting the stopping criterions;
5. Obtaining $H^*$ via determining sample-cluster membership with parameter $\tau$;
6. Generating cluster set $E_K$ from the refined matrix $H^*$.

**Figure 4 The algorithm of Meta-EC for microbial community pattern detection**.

our study. Then we conducted experimental comparison between Meta-EC and base clustering approaches, and comparison between constructed consensus network and original metagenomic similarity network. Finally, from clustering results, we investigated how human microbial community was influenced by body habitat and host gender.

**Evaluation metrics**
In this work, we evaluated the effectiveness of clustering algorithms by observing how well detected clusters corresponded to the sampling information of habitats and genders (six meta-classes, refer to **subsection Terminology** for details). Since the true number of cluster patterns for habitat and gender was unknown, and there were no literature references to clearly mention how to determine the number of cluster patterns in either body habitat or host gender, we empirically defined reference clusters based on six **meta-classes**. Assuming that metagenomic samples with identical **meta-classes** were likely to have similar microbial structures [13], we bring the metagenomic samples with identical **meta-classes** into one reference cluster. Typically, the quality of the predicted clusters could be evaluated by following three quantity measures, *f*-measure [33], PR metrics and F-score, which could measure how well the detected clusters corresponded to reference clusters.

Among these three measures, *f*-measure which was the harmonic mean of Precision and Recall, aimed at assessing how well the detected clusters matched reference ones at cluster level (Precision measured what fraction of the detected clusters were matched with reference ones and Recall measured what fraction of reference clusters were matched to detected clusters). PR-based metric took into account the overlap between detected and reference clusters. F-score focused on measuring whether samples within identical habitats were grouped together in the detected clusters. The value of each measure varied from 0 to 1, and the higher value indicated better match. For more details of $f$-measure, PR metrics and F-score, please refer to Additional file 1: Section 2. And the parameter setting in the experiments is introduced in Additional file 1: Section 3.

### Evaluation of clustering results generated by Meta-EC algorithm

In this subsection, to evaluate the performance of Meta-EC algorithm, we presented performance comparison of proposed Meta-EC algorithm with base clustering approaches and comparison of constructed consensus matrix with original microbial similarity matrix.

**Comparison against four base clustering approaches:** To evaluate the performance of ensemble clustering approach, the accuracy of the clustering results derived from our proposed approach was compared with the ones derived from these base clustering algorithms. Figure 5 illustrated the performance of different clustering algorithms in terms of three metrics (PR, $f$-measure and F-score) with respect to the reference clusters. From Figure 5, we could observe that our ensemble-based approach had competitive performance compared with the base clustering algorithms as regard to all three measures. Among the base clustering algorithms, K-means with cluster number set to 6 had better performance in terms of PR, while K-means and Density-based clustering with cluster number set to 6 had better performance in terms of $f$-measure, and Hierarchical clustering with cluster number set to 9 and 10 had comparable performance with K-means and Density-based clustering with cluster number set to 6 in terms of F-score. But none of them could have superior performance than others as regard to all three measures. However, our ensemble-based approach obtained the best performance in terms of all the three measures. This may be owing to the fast that ensemble-based approach could make use of clusters derived from different base clustering algorithms and extract more reliable results. In addition, we conducted sensitivity study of phylogenetic structure similarity on microbial network. We ran algorithm Meta-EC with threshold value of metagenomic similarity in matrix tuning from 0.7 to 0.9 with 0.1 as step size, the results in Additional file 2: **Figure S4** showed Meta-EC outperformed other state-of-art clustering techniques in the wide range of edge threshold, indicating that our algorithm is robust and insensitive to the similarity network noisy and data coverage. In addition, we have compared the computational time with base clustering approaches in Table 3 and results show that Meta-EC exactly spend more time than that by K-mean and Hierarchical clustering but less than EM clustering, so the total time cost of MetaEC is the sum of all base clustering algorithms plus 72.8 seconds. With rapid development of computational capability, we could improve the time efficiency on large amount of operations.

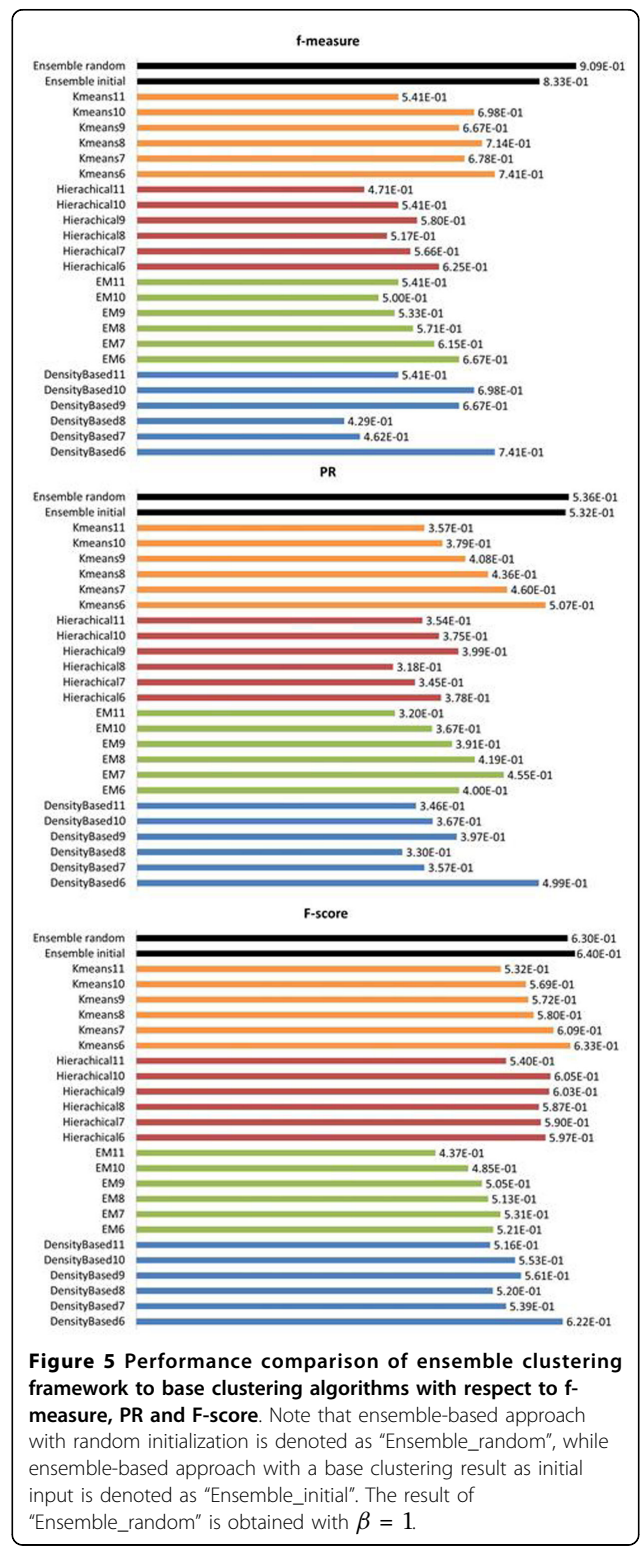

**Figure 5 Performance comparison of ensemble clustering framework to base clustering algorithms with respect to f-measure, PR and F-score.** Note that ensemble-based approach with random initialization is denoted as "Ensemble_random", while ensemble-based approach with a base clustering result as initial input is denoted as "Ensemble_initial". The result of "Ensemble_random" is obtained with $\beta = 1$.

**Table 3 Comparison with bases clustering approaches on computational time**

| Method  | EM     | K-Means | Hierarchical | Density-based | Meta-EC |
|---------|--------|---------|--------------|---------------|---------|
| **Time(S)** | 221.17 | 11.57   | 55.04        | 12.24         | 72.8    |

**Comparison of constructed consensus network with original similarity network:** To demonstrate the benefits of combining different base clustering results, we applied symmetric NMF on original metagenomic similarity network and evaluated its performance. To be fair, the results of symmetric NMF on original metagenomic similarity network were obtained over the best tuned parameter. The comparison of the two tested similarity network is present in Figure 6 as regard to F-measure, PR and f-measure.

The results in Figure 6 showed that applying symmetric NMF on consensus matrix achieved better performance than that on the original similarity network. These results demonstrated the benefits of combining different base clustering results. If the similarity matrix was well constructed (each element reflected the co-cluster similarity), the factorization of the similarity matrix would generate a clustering assignment matrix that could well capture the cluster structure inherent in the network representation. However, the original network weighted the interaction via measuring the phylogenetic structure of samples. In this way, metagenomic samples with higher phylogenetic similarity were more likely to be involved in one cluster. If the actual microbial pattern was uncorrelated with phylogenetic similarity, the community detected by symmetric NMF may be unreliable. In ensemble clustering framework, we generated a consensus matrix that integrated the clustering results derived from different clustering algorithms. Each element in consensus matrix indicated the frequency of the corresponding sample pair being clustered together in these base clustering results. Thus, applying symmetric NMF on consensus matrix could take into consideration the co-cluster strength of multiple clustering patterns and output a more comprehensive and robust result.

### Interpretation of Microbial community patterns on human body habitats based on clustering results

Recall that metagenomic samples were clustered in terms of co-occurrence frequency in base clustering results. Hence the final output clusters assembled samples to represent unique microbial patterns that are the consensus from base clustering approaches. Next, from the clustering results, we infer how microbial pattern was influenced by body habitats and host genders.

**Structural variation across body habitats:** Through analyzing the enrichment of body habitat and host gender over six predicted clusters, the results in Figure 7 revealed a stronger coherence by body habitat than host gender. These clusters dominated by particular body habitats inferred that these body habitats harboured distinctive microbial patterns, which was also observed in base clustering results in Additional file 2: Figure S3. Although four base clustering algorithms generate clustering patterns with different criterions, most clusters in Additional file 2: Table S2 were enriched with particular habitats.

Meanwhile, we observed that microbial communities at different body habitat exhibited different degree of compositional structure variation. Figure 7 showed that microbial structure remained relatively stable in oral cavity, compared with diverse microbial structures harboured in skin. It was biologically reasonable to detect diverse patterns on skin, since there were quite different places where skin microbial communities could be sampled.

Different extend of habitat structural variation were also observed in base clustering results. In Additional file 2: Figure S2, gut and oral cavity microbial community patterns were only fit with one clustering criterion, gut consistent with K-means and oral cavity with hierarchical clustering. Contrary to gut and oral cavity, skin-enriched cluster could be recognized by four clustering criteria in all experimental settings, inferring skin samples have many cluster patterns with diverse microbial structures.

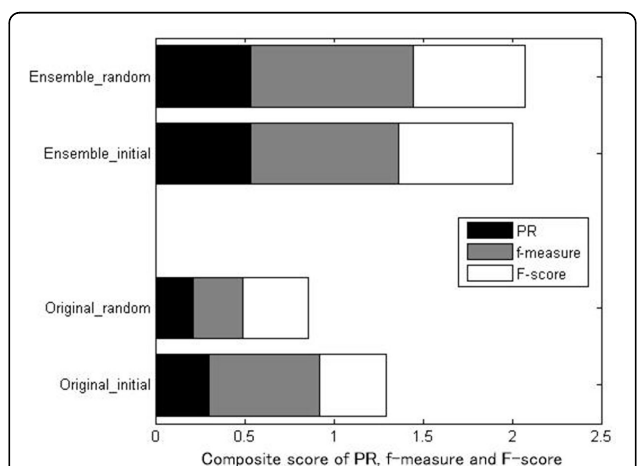

**Figure 6 Performance comparison of Bayesian NMF based clustering algorithm applied on ensemble clustering similarity network and original microbial similarity network.** Additive values of three measures are present for each data source. For random initialization case, the value of $\beta$ is set to 1 and the result corresponds to "Original_random". We also choose the base clustering results which presents the best performance as the initial input of symmetric NMF and the result corresponds to "Original_initial".

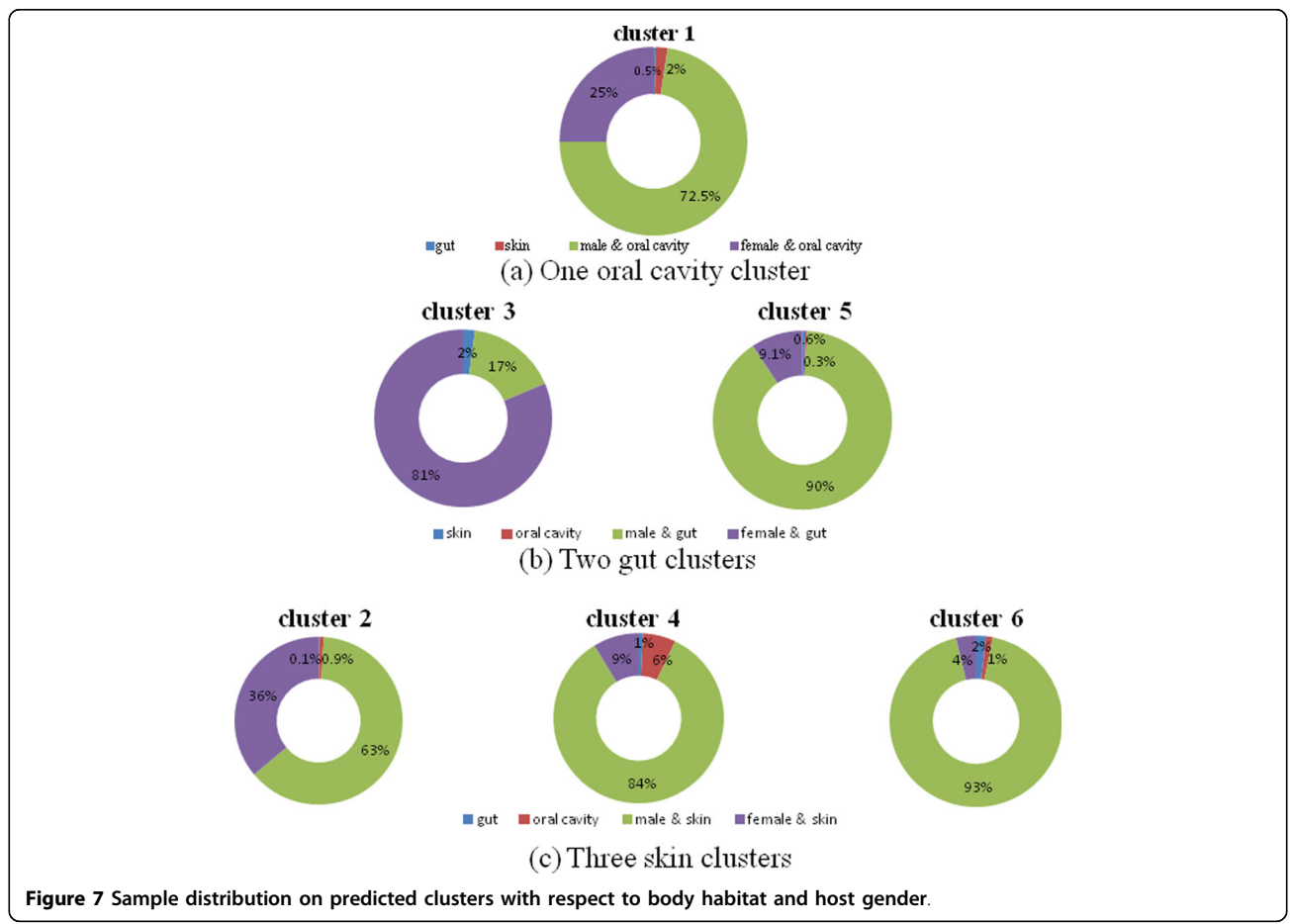

**Figure 7 Sample distribution on predicted clusters with respect to body habitat and host gender**.

Note that the proposed Meta-EC generates a more comprehensive community patterns with respect to meta-data since our result is an agreement by consensus of multiple base clustering approaches. For example, compared to Hierarchical and EM clustering results in Additional file 2: **Figure S3** that only capture male-gut cluster, ensemble clustering is able to uncover female-gut specific clusters (shown in Figure 7), indicating that Meta-EC could reveal degree of structural variation over body habitat more comprehensively than base clustering results.

**Structural variation across host gender:** We further assessed microbial structure variation with respect to host gender. Meta-Storm [27] was used to measure similarity of two metagenomic samples. The results in Figure 8 indicated that over all habitats, variation was significantly less within same gender samples than between opposite gender samples. However, these habitats perform different degree of structural variation with respect to host gender. Oral cavity microbiome exhibited a stable structure both among same and opposite gender individuals (both above 92% phylogenetic structure similarity). And skin communities had no unique structural variation patterns regarding to host gender. Gut community structure was highly variable between samples from opposite gender hosts (less than 90% similarity value for opposite gender samples of gut cluster 3), but exhibited strong coherence to same gender hosts. On the other hand, the enrichment study in Figure 7 showed that two gut clusters were distinct with host gender, indicating that opposite sexual individuals may exhibit a distinct microbial composition in gut.

**Microbial interconnection over habitats:** Although microbial communities reflected unique structures (distributions) over body habitats, the interconnected microbial components among the body habitats were still observed in the clustering results. For example, cluster 1 in Figure 7 contained 10 skin samples that shared similar microbial compositions with oral cavity communities, while skin cluster 2, 4 and 6 harboured 6, 15 and 2 oral cavity samples respectively. Since skin microbial pattern was closely associated with external environment [34] and oral cavity was an open system where microbiome from external environment was imported by breathing, eating food and drinking water [35], oral cavity and skin would respond to outside environmental conditions, and gradually evolve similar microbiomes.

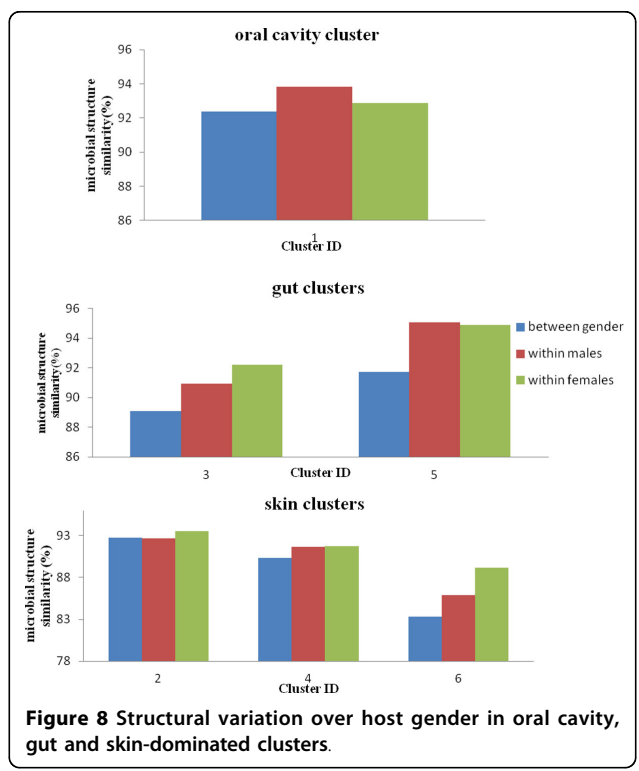

**Figure 8 Structural variation over host gender in oral cavity, gut and skin-dominated clusters**.

## Conclusions and discussions

The human microbiomes are microbiomes that are hosted in gut, oral mucosa and multi-layer of skin etc. These organisms perform ecosystem-level functions that are useful for human host to maintain healthy, yet detailed factors that attribute the microbial community structures in human body habitats and host gender remain poorly conceptualized. To fully understand the roles of human microbiome in disease and health, prior studies focus on particular body habitats of health individuals with specific clustering approaches, based on the assumption that metagenomic samples of same body habitats would develop similar microbial structure patterns. However, human habitats are not isolated; they are interacted and correlated to form an integrated and complex system. And identified structures might be unsuccessful due to noisy sample similarity and specific topological structure within metagenomic network. Hence, single clustering algorithm rarely achieves optimal outcome. To uncover a global and comprehensive landscape of human microbiome, we perform an ensemble clustering framework **Meta-EC** on large-scale metagenomic samples. In this study, our proposed Meta-EC algorithm has four main advantages on microbial pattern detection: (1) Meta-EC could effectively identify more reliable microbial communities via integrating many base clustering results, (2) As regard to the modularity of microbial communities, defined as the clustering of microbial communities (modularity) according to the effects of their related environments or treatments (meta-data), the consensus clustering network is much clearer at showing such modularity property (such as how environments (or meta-data) shape microbial communities in body habitat, which is critical to healthcare and diognosis) than the original metagenomic similarity network [25], (3) Ensemble framework is robust for the coverage of metagenomic similarity network (as shown in Additional file 2: **Figure S4**), and (4) Compared to base clustering results in Additional file 2: **Figure S3**, Meta-EC algorithm could reveal the spatial and gender patterns of microbiome (as shown in Figure 7) more comprehensively, as the ensemble clustering result is a general agreement by multiple base clustering approaches.

Nevertheless, it should be acknowledged that the performance of our algorithm depends on the base clustering results and quality of original metagenomic similarity network. If all these base results were generated by poor clustering algorithms, the ensemble outputs would be far from real microbial community similarity patterns. If the original similarity network is unreliable to capture the modularity of metagenomic samples, none of clustering approaches could work. To address this problem, we have to integrate more base clustering approaches with diverse optimization criterions and pattern assumptions, to reduce the bias generated by base approaches. We assume these algorithms can capture a wide variety of clustering patterns in similarity network to alleviate the effect of unreliable clustering results. On the other hand, the proposed NMF based mode, which could be used in association study of bioinformatics domain [36-39], is a more complex method to implement, and convergence could be slow, as shown in Table 3. With rapid development of computational capability, we could improve the time efficiency on large amount of operations. And the nonnegative constraints on cluster indicator matrix $H$ may be an insufficient condition for achieving sparseness in some cases [20]. Then one may set appropriate thresholds to enforce sparseness. In summary, Meta-EC is an ensemble clustering framework for large-scale metagenomic data analysis and microbial community pattern detection. In the future, NMF based model could be exploited to offer potential applications on bipartite model of drug-target association [40] and disease gene prediction [41].

## Availability
The data sets and supporting experimental results of this article are available for download from http://datam.i2r.a-star.edu.sg/MetaEC.

## Additional material

Additional file 1: Experimental Design. The file show the experimental design in this paper, including: (1) introductory of four base clustering approaches; (2) evaluation of microbial clusters; (3) parameter setting.

Additional file 2: Supplementary Material. The file presents several figures, tables and additional experimental results mentioned in this paper, including: (1) the efficiency of GPU-meta-storm algorithm; (2) evaluation of four base clustering results; (3) sensitivity study of phylogenetic structure similarity on microbial network.

**Competing interests**
The authors declare that they have no competing interests.

**Authors' contributions**
Conceptualed and designed the method and drafted manuscript: PY KN. Responsible for the implementation: PY XS LOY. Provided raw data: KN XS. Participated in discussion and improved the method as well as revised the draft: XS LOY H-NC X-LL. Read and approved the manuscript: PY XS LOY H-NC X-LL KN.

**Acknowledgements**
This work is supported in part by Chinese Academy of Sciences' e-Science grant INFO-115-D01-Z006, Ministry of Science and Technology's high-tech (863) grant 2009AA02Z310 and 2014AA21502, as well as National Science Foundation of China grant 61103167 and 31271410.

**Declarations**
Publication costs for this article were partially funded by Chinese Academy of Sciences' e-Science grant INFO-115-D01-Z006, Ministry of Science and Technology's high-tech (863) grant 2009AA02Z310 and 2014AA21502, as well as National Science Foundation of China grant 61103167 and 31271410 and by the Institute for Infocomm Research, Agency for Science, Technology & Research (A*STAR), Singapore.
This article has been published as part of *BMC Systems Biology* Volume 8 Supplement 4, 2014: Thirteenth International Conference on Bioinformatics (InCoB2014): Systems Biology. The full contents of the supplement are available online at http://www.biomedcentral.com/bmcsystbiol/supplements/8/S4.

**Authors' details**
[1]Institute for Infocomm Research, Agency for Science, Technology & Research (A*STAR), Singapore, 138632, Singapore. [2]Computational Biology Group of Single Cell Center, Shandong Key Laboratory of Energy Genetics and CAS Key Laboratory of Biofuels, Qingdao Institute of Bioenergy and Bioprocess Technology, Chinese Academy of Science, Qingdao 266101, China. [3]Center for Computer Vision and Department of Mathematics, Sun Yat-Sen University, Guangzhou, 510275, China.

Published: 8 December 2014

**References**
1. Wilson M: **Bacteriology of humans: an ecological perspective.** *John Wiley & Sons* 2009.
2. Dethlefsen L, McFall-Ngai M, Relman DA: **An ecological and evolutionary perspective on human-microbe mutualism and disease.** *Nature* 2007, **449(7164)**:811-818.
3. Turnbaugh PJ, Ley RE, Hamady M, Fraser-Liggett CM, Knight R, Gordon JI: **The human microbiome project: exploring the microbial part of ourselves in a changing world.** *Nature* 2007, **449(7164)**:804-810.
4. Lederberg J: **Infectious history.** *Science* 2000, **288(5464)**:287-293.
5. Eckburg PB, Bik EM, Bernstein CN, Purdom E, Dethlefsen L, Sargent M, Gill SR, Nelson KE, Relman DA: **Diversity of the human intestinal microbial flora.** *Science* 2005, **308(5728)**:1635-1638.
6. Fierer N, Hamady M, Lauber CL, Knight R: **The influence of sex, handedness, and washing on the diversity of hand surface bacteria.** *Proceedings of the National Academy of Sciences* 2008, **46**:17994-17999.
7. Aas JA, Paster BJ, Stokes LN, Olsen I, Dewhirst FE: **Defining the normal bacterial flora of the oral cavity.** *Journal of Clinical Microbiology* 2005, **43(11)**:5721-5732.
8. Nasidze I, Quinque D, Li J, Li M, Tang K, Stoneking M: **Comparative analysis of human saliva microbiome diversity by barcoded pyrosequencing and cloning approaches.** *Analytical biochemistry* 2009, **391(1)**:64-68.
9. Turnbaugh PJ, Hamady M, Yatsunenko T, Cantarel BL, Duncan A, Ley RE, Sogin ML, Jones WJ, Roe BA, Affourtit JP, et al: **A core gut microbiome in obese and lean twins.** *Nature* 2009, **7228**:480-484.
10. Grice EA, Kong HH, Conlan S, Deming CB, Davis J, Young AC, NISC Comparative Sequencing Program, Bouffard GG, Blakesley RW, Murray PR, et al: **Topographical and temporal diversity of the human skin microbiome.** *Science* 2009, **324(5931)**:1190-1192.
11. Bik EM, Long CD, Armitage GC, Loomer P, Emerson J, Mongodin EF, Nelson KE, Gill SR, Fraser-Liggett CM, Relman DA: **Bacterial diversity in the oral cavity of 10 healthy individuals.** *The ISME journal* 2010, **4(8)**:962-974.
12. Mitreva M: **Structure, function and diversity of the healthy human microbiome.** *Nature* 2012, **486**:207-214.
13. Costello EK, Lauber CL, Hamady M, Fierer N, Gordon JI, Knight R: **Bacterial community variation in human body habitats across space and time.** *Science* 2009, **5960**:1694-1697.
14. Lozupone C, Hamady M, Knight R: **UniFrac-an online tool for comparing microbial community diversity in a phylogenetic context.** *BMC Bioinformatics* 2006, **7**:371.
15. Kent AD, Yannarell AC, Rusak JA, Triplett EW, McMahon KD: **Synchrony in aquatic microbial community dynamics.** *The ISME journal* 2007, **1(1)**:38-47.
16. Zinger L, Coissac E, Choler P, Geremia RA: **Assessment of microbial communities by graph partitioning in a study of soil fungi in two alpine meadows.** *Applied and environmental microbiology* 2009, **75(18)**:5863-5870.
17. Lloyd SP: **Least squares quantization in PCM.** *IEEE Transactions on Information Theory* 1982, **28**:129-137.
18. Szekely GJ, Rizzo ML: **Hierarchical clustering via Joint Between-Within Distances: Extending Ward's Minimum Variance Method.** *Journal of classification* 2005, **22(2)**:151-183.
19. Moon TK: **The expectation-maximization algorithm.** *IEEE Signal processing magazine* 1996, **13(6)**:47-60.
20. Devarajan K: **Nonnegative matrix factorization: an analytical and interpretive tool in computational biology.** *PLoS computational biology* 2008, **4(7)**:e1000029.
21. Qi Q, Zhao Y, Li M, Simon R: **Non-negative matrix factorization of gene expression profiles: a plug-in for BRB-ArrayTools.** *Bioinformatics* 2009, **25(4)**:545-547.
22. Zhang S, Li Q, Liu J, Zhou XJ: **A novel computational framework for simultaneous integration of multiple types of genomic data to identify microRNA-gene regulatory modules.** *Bioinformatics* 2011, **27(13)**:i401-i409.
23. Ou-Yang L, Dai DQ, Zhang XF: **Protein complex detection via weighted ensemble clustering based on Bayesian nonnegative matrix factorization.** *PloS One* 2013, **8(5)**:e62158.
24. Lancichinetti A, Fortunato S: **Consensus clustering in complex networks.** *Scientific reports* 2012, **2**.
25. Kuang D, Park H, Ding CH: **Symmetric Nonnegative Matrix Factorization for Graph Clustering.** *SDM* 2012, **12**:106-117.
26. Caporaso JG, Lauber CL, Costello EK, Berg-Lyons D, Gonzalez A, Stombaugh J, Knights D, Gajer P, Ravel J, Fierer N, et al: **Moving pictures of the human microbiome.** *Genome Biol* 2011, **12**:R50.
27. Su X, Xu J, Ning K: **Meta-Storms: Efficient Search for Similar Microbial Communities Based on a Novel Indexing Scheme and Similarity Score for Metagenomic Data.** *Bioinformatics* 2012, **28(19)**:2493-2501.
28. Kullback S: **Letter to the Editor: The Kullback-Leibler distance.** *The American Statistician* 1987, **41(4)**:340-341.
29. Psorakis I, Roberts S, Sheldon B: **Soft partitioning in networks via bayesian non-negative matrix factorization.** *Adv Neural Inf Process Syst* 2010.
30. Tan VY, Févotte C: **Automatic relevance determination in nonnegative matrix factorization.** *In SPARS'09-Signal Processing with Adaptive Sparse Structured Representations* 2009.

31. Seung D, Lee L: **Algorithms for non-negative matrix factorization.** *Advances in neural information processing systems* 2001, **13**:556-562.
32. Greene D, Cagney G, Krogan N, Cunningham P: **Ensemble non-negative matrix factorization methods for clustering protein-protein interactions.** *Bioinformatics* 2008, **24**(15):1722-1728.
33. Manning CD, Raghavan P, Schütze H: **Introduction to information retrieval.** *Cambridge university press* 2008, **1**:6.
34. McGuire AL, Colgrove J, Whitney SN, Diaz CM, Bustillos D, Versalovic J: **Ethical, legal, and social considerations in conducting the Human Microbiome Project.** *Genome Research* 2008, **18**(12):1861-1864.
35. Dewhirst FE, Chen T, Izard J, Paster BJ, Tanner AC, Yu WH, Wade WG: **The human oral microbiome.** *Journal of bacteriology* 2010, **192**:(19): 5002-5017.
36. Yang P, Li X, Mei JP, Kwoh CK, Ng SK: **Positive-unlabeled learning for disease gene identification.** *Bioinformatics* 2012, **28**(20):2640-2647.
37. Mei JP, Kwoh CK, Yang P, Li X, Zheng J: **Drug-target interaction prediction by learning from local information and neighbors.** *Bioinformatics* 2013, **29**(2):238-245.
38. Yang P, Li X, Wu M, Kwoh CK, Ng SK: **Inferring gene-phenotype association via global protein complex network propagation.** *PloS One* 2011, **6**(7):e21502.
39. Zheng X, Ding H, Mamitsuka H, Zhu S: **Collaborative matrix factorization with multiple similarities for predicting drug-target interactions.** *In 19th ACM SIGKDD international conference on Knowledge discovery and data mining* 2013, 1025-1033.
40. Mei JP, Kwoh CK, Yang P, Li X, Zheng J: **Globalized bipartite local model for drug-target interaction prediction.** *Proceedings of the 11th International Workshop on Data Mining in Bioinformatics* 2012, 8-14.
41. Yang P, Li X, Chua HN, Kwoh CK, Ng SK: **Ensemble Positive Unlabeled Learning for Disease Gene Identification.** *PloS one* 2014, **9**(5):e97079.